**Measurement and comparison of distributional shift with applications to ecology, economics, and image analysis**

Kenneth J. Locey[1*], Brian D. Stein[1]

[1] Center for Quality, Safety and Value Analytics, Rush University Medical Center, Chicago, Illinois, 60612, USA.

[*] Corresponding author. email: Kenneth_J_Locey@rush.edu

**Abstract**

The concentration of a distribution toward a lower bound is a conceptually simple property that closely relates to concepts of rarity and poverty, but that lacks a global descriptive statistic. We term this property 'shift' and define it as the distance of a central tendency from an upper bound, expressed as a proportion of a finite range. We derive a flexible, low complexity measure of shift and demonstrate its properties, its use with theoretical distributions, and its relation to skewness. We then use shift as the basis for a directional difference measure and as the basis for a formal distance metric that closely approximates the behavior of metrics having greater complexity (e.g., Wasserstein distance). Using simulated datasets and comparisons to system-specific measures, we demonstrate shift as a measure of species rarity and as a measure of poverty. We then apply our shift statistics to the analysis of image data. The shift statistics presented have a high degree of potential use across disciplines.

## 1 Introduction

Distributions of wealth, size, and abundance are generally right-skewed, heavy-tailed, and take similar forms across disparate systems [41]. Numerous theories, models, and empirical laws have been proposed to explain the ubiquity of these highly uneven distributions as emergent properties of complex systems, as equilibrium outcomes of stochastic processes, and as statistical inevitabilities of numerical constraints [35, 41, 42]. Likewise, many descriptive statistics have been developed to measure intensively studied properties of these distributions, e.g., diversity, evenness, inequality [24, 39, 53]. And yet, a conspicuous feature of these distributions that closely relates to other intensively studied empirical properties remains undefined, i.e., the concentration of a distribution toward a lower bound.

In ecology, rarity commonly pertains to the concentration of species toward the lowest abundances and is often measured via skewness or thresholds of abundance [33, 37-39, 51]. However, skewness is a property of asymmetry and not concentration at location and, for distributions of $N$ individuals among $S$ species, skewness is numerically constrained by $N$ and $S$ [35]. Consequently, measures of rarity based on skewness can be misleading and dependent on the values of general state variables. In contrast, measures of rarity based on abundance thresholds are intuitive and reflect concentration at location. Examples are the percentage of species represented by one individual ($R_1$) and the percentage of species having relative abundances below 1% ($R_{1\%}$) [33, 39]. However, such measures are fragile [39]. For example, when using $R_{1\%}$, no species can be rare if $N$ is less than 101. Simply put, if $x \leq 100/N$ then $R_{x\%} = 0$. By the same measure, at least 201 species must be rare if $N = 10^6$ and $S = 300$. Yet, for the same values of $N$ and $S$, few if any species would be considered rare when using $R_1$, as the average species abundance ($N/S$) would be nearly 3333.

The concentration of a distribution toward a lower bound also relates to poverty, i.e., the concentration of a population toward its lowest incomes [17, 24]. Measures of poverty invariably rely on poverty lines (Table 1) [17, 24, 44]. In the US, poverty lines are based on estimated costs of adequate food intake for families of specific sizes [24, 48]. As such, poverty lines have long been considered arbitrary and over-simplified [24]. Though poverty lines have practical uses, they also create false divides among the impoverished [24] and are vulnerable to manipulation [2, 55]. Unlike measures of poverty, measures of economic inequality are typically based on general properties of statistical dispersion [24, 53]. This discrepancy begs the question of whether measures of poverty could also be based on a general property.

**Table 1.** Five common poverty measures and the Gini coefficient of inequality as defined in Haughton and Khandker [24]

| Measure | Equation | Description |
|---|---|---|
| Poverty rate, i.e., head count ratio | $P_0 = \frac{N_P}{N}$ | The number of impoverished people, families, etc. ($N_P$) divided by the number of people, families, etc. ($N$). |
| Poverty Gap Index | $P_1 = \frac{1}{N}\sum_{i=1}^{N}\frac{G_i}{z}$ | Normalized sum of ratios between individual poverty gaps ($G_i$) and the poverty line ($z$), where $G_i = z - y_i$ and $y_i$ is the income of the $i^{th}$ individual, family, etc. |
| Sen Index | $P_S = P_0\left(1 - (1 - G^P)\frac{\mu^P}{z}\right)$ | A combination of poverty rate ($P_0$) and Gini's inequality index ($G$), where $G^P$ is inequality among the poor and $\mu^P$ is the mean income among the poor. |
| Sen-Shorrocks-Thon Index | $P_{SST} = P_0 P_1^P\left(1 + \hat{G}^P\right)$ | The product of poverty rate, the poverty gap index (only including the poor), and 1 plus the Gini coefficient of the poverty gap ratios for the entire population. |
| Watts Index | $W = \frac{1}{N}\sum_{i=1}^{P} ln\left(\frac{z}{y_i}\right)$ | Normalized sum of logarithmically transformed ratios of the poverty line ($z$) to individual income ($y_i$) for all individuals qualifying as poor ($P$). $N$ is the total number of individual, households, etc. |
| Gini coefficient | $1 - \sum_{i=1}^{N}(x_i - x_{i-1})\,(y_i + y_{i-1})$ | Income inequality as 1 minus the sum of weighted differences between consecutive cumulative income shares, where ($x_i - x_{i-1}$) represents change in the cumulative income distribution and ($y_i + y_{i-1}$) represents the cumulative population share. $N$ is the total number of individual, households, etc. |

The concentration of a distribution toward a lower (or upper) bound also relates to the direction and magnitude by which one distribution is shifted to lower or higher values relative to another. For example, shift functions visualise directional differences between distributions, but do not precisely quantify them [12, 46]. In contrast, some distance metrics quantify differences between distributions in ways that relate to shift, but are non-directional (Table 2) [3, 4, 15, 20, 31]. Yet, distance metrics are used across sciences [e.g., 5, 29, 36] and are often embedded into the training and evaluation of deep learning models [e.g., 4, 22, 27, 47]. Distance metrics are particularly useful in analyses of image data based on comparisons of histograms of hue, saturation, intensity, and other visual properties [7, 11, 14, 18, 27]. However, established metrics would not distinguish whether one image is shifted to higher or lower values of a particular visual property relative to another image.

**Table 2.** Select distance metrics commonly used to compare frequency distributions. Equations and description modified from Deza and Deza, 2014 [15]

| Measure | Equation | Description |
|---|---|---|
| Wasserstein Distance | $min_{\gamma \in \Pi(P,Q)} \sum_{i,j} \gamma(i \cdot j) \cdot d(i,j)$ | Measures dissimilarity between cumulative probability distributions by quantifying the minimum "work" required to transform one into another. The equation originates from optimal transport theory. |
| Kolmogorov-Smirnov Distance | $sup_x \lvert F_1(x) - F_2(x) \rvert$ | The maximum absolute difference between cumulative distribution functions. |
| Energy Distance | $D(F_1, F_2) = 2(E\lvert X-Y \rvert - E\lvert X-X' \rvert - E\lvert Y-Y' \rvert)^{1/2}$ | For 1D distributions, measures dissimilarity between two cumulative distributions using the sum of pairwise squared distances between samples within each distribution and between samples from different distributions. |
| Cramér Von Mises Distance | $\int_{-\infty}^{\infty} [F_n(x) - F^*(x)]^2 dF^*(x)$ | Squared differences between empirical cumulative distribution functions integrated across all values. |
| Hellinger Distance | $\frac{1}{2}\int_{-\infty}^{\infty} \left(\sqrt{f_1(dx)} - \sqrt{f_2(dx)}\right)^2$ | Distance between two distributions computed as the integral (or sum) of the squared differences between the square roots of their probability density functions. |
| Total Variation Distance | $\frac{1}{2}\int_{-\infty}^{\infty} (\lvert f_1(x) - f_2(x) \rvert)\, dx$ | Largest difference in probabilities assigned by two distributions measured as half the integral of the absolute difference between probability functions ($f_1$, $f_2$). |

In this work, we define the shift statistic, $\mathcal{S}$, for discrete and continuous distributions of probabilities or frequencies as the location of a defined central tendency relative an upper bound, expressed as a proportion of a finite interval. We explain the properties of $\mathcal{S}$ and demonstrate its use with theoretical distributions and its relation to skewness. We then use $\mathcal{S}$ as the basis for a directional comparative measure, $\Delta\mathcal{S}$, and demonstrate $|\Delta\mathcal{S}|$ as a proper distance metric. We demonstrate the application of $\mathcal{S}$ and $\Delta\mathcal{S}$ to the measurement of poverty and species rarity, and demonstrate the use of $\mathcal{S}$, $\Delta\mathcal{S}$, and $|\Delta\mathcal{S}|$ in the analysis of image data.

## 2 Shift as a descriptive statistic

### 2.1 Definition

Let $D$ be a discrete or continuous distribution of probabilities or frequencies. Suppose $[L, U]$ is a closed interval with $L < U$ containing the support of $D$ (for a probability distribution) or the range of observed data (for a frequency distribution):

$$\text{supp}(D) \subseteq [L, U] \quad \text{or} \quad \text{observed values} \subseteq [L, U],$$

Note that $[L, U]$ need not be the minimal interval containing the support or observed data, but may extend beyond these to include regions of zero probability or zero frequency, as when $[L, U]$ represents a fixed reference domain.

We define the shift statistic, denoted $\mathcal{S}$, as the location of a defined central tendency of $D$ (e.g., mean, median, mode), denoted $C$, relative to the upper bound of $[L, U]$, expressed as a proportion of the interval's width:

$$\mathcal{S} = \frac{U - C}{U - L}, \tag{1}$$

where $\mathcal{S} \in [0, 1]$, since $L < U$.

The values of $\mathcal{S}$ has a straightforward interpretation. If $\mathcal{S} = 0$, then $C$ lies at the upper bound $U$. If $\mathcal{S} = 1$, then $C$ lies at the lower bound $L$. Otherwise, $0 < \mathcal{S} < 1$ and $C$ is shifted from $U$ by a proportion of the interval equal to $\mathcal{S}$. Since the interval $[L, U]$ need not be the minimal interval containing $D$, $\mathcal{S}$ is defined when $D$ is degenerate. For example, let $D_i$ be a discrete probability distribution that is degenerate at the value $i \in \{1, 2, 3\}$. Taking $D_i$ within $[L, U]$, where $L = 1$ and $U = 3$:

$$D_{i=1} = [1, 0, 0],\ C = 1,\ \mathcal{S} = (3 - 1)/(3 - 1) = 1.$$
$$D_{i=2} = [0, 1, 0],\ C = 2,\ \mathcal{S} = (3 - 2)/(3 - 1) = ½,$$
$$D_{i=3} = [0, 0, 1],\ C = 3,\ \mathcal{S} = (3 - 3)/(3 - 1) = 0.$$

To our knowledge, $\mathcal{S}$ is not redundant with other descriptive statistics. However, the form of $\mathcal{S}$ resembles the data scaling technique of mean normalization [29]:

$$x' = (x - \bar{x}) / (\max(x) - \min(x)),$$

where $\bar{x}$, $\min(x)$, and $\max(x)$ denote the mean, minimum, and maximum values of the data vector $x$. Mean normalization evaluated at $\max(x)$ yields $\mathcal{S}$ for $C = \bar{x}$, $L = \min(x)$, $U = \max(x)$. Though mean normalization offers a specific instance of $\mathcal{S}$, it does not generalize to a flexible statistic. In contrast, the definition of $\mathcal{S}$ allows $C$ to be the expected value of a theoretical distribution having bounded support, e.g., Beta, Binomial, Hypergeometric. Consider the Beta distribution:

$$f(x|\alpha, \beta) = \frac{x^{\alpha-1}(1-x)^{\beta-1}}{B(\alpha, \beta)},\ 0 \leq x \leq 1,$$

where $\alpha > 0$, $\beta > 0$ are shape parameters, $B$ is the Beta function, and $\mathbb{E}[X] = \alpha/(\alpha + \beta)$ is the expected value. Taking $C = \mathbb{E}[X]$, $L = 0$, $U = 1$, $\mathcal{S}$ within the domain of the distribution is:

$$\mathcal{S} = \frac{U - \mathbb{E}[X]}{U - L} = 1 - \mathbb{E}[X] = \frac{\beta}{\alpha + \beta}. \tag{2}$$

Likewise for the Binomial distribution, where the possible number of successes in $n$ independent trials ranges from 0 to $n$, each with success probability $p$, and where $\mathbb{E}[X] = np$ is the expected number of successes. Taking $C = \mathbb{E}[X]$, $L = 0$, $U = n$, shift of the Binomial distribution within its support becomes:

$$\mathcal{S} = \frac{U - \mathbb{E}[X]}{U - L} = \frac{n - np}{n} = 1 - p.$$

By definition, $\mathcal{S}$ is not calculable for theoretical distributions with unbounded support, e.g., Gamma, Negative binomial, Poisson. However, such distributions are often truncated and re-normalized within finite bounds [21, 40]. Letting $L$ and $U$ be the bounds of truncation, the interval $[L, U]$ may capture an arbitrarily large proportion $m \in (0, 1)$ of the total probability mass. Consider the Poisson distribution [28]:

$$P(X = k) = \frac{\lambda^k e^{-\lambda}}{k!}, \qquad k \in \mathbb{N}_0,$$

where $\lambda > 0$ is the expected count and $k$ is the observed count from 0 to $+\infty$. Using the quantile function to define equal-tailed bounds enclosing mass $m$, yields:

$$L = Q_\lambda\left(\frac{1 - m}{2}\right), \quad U = Q_\lambda\left(1 - \frac{1 - m}{2}\right).$$

Optimizing $m$ to preserve the moments of the unbounded Poisson distribution reveals that, as $\lambda$ increases across orders of magnitude, $\mathcal{S}$ approaches ½ while the moment coefficient of skewness ($\gamma_1$) approaches 0, suggest that $\mathcal{S} \to$ ½, $\gamma_1 \to 0$ as $\lambda \to +\infty$ (Fig 1). This relationship is consistent with how skewness inversely relates to the square root of $\lambda$ in the canonical Poisson distribution, $\gamma_1 = 1/\sqrt{\lambda}$ (Fig 1).

**Figure 1.** Analysis of a truncated, renormalized Poisson distribution wherein truncation limits were optimized to preserve the moments of the unbounded distribution. In each subplot, 'Approx.' pertains to values calculated from the truncated distribution. **A:** Skewness ($\gamma_1$) of the unbounded Poisson distribution vs. the mean ($\lambda$). **B:** Approximated value of $\lambda$ versus the true value. **C:** Approximated value of $\gamma_1$ versus the true value. **D:** Approximated variance ($\sigma^2$) versus the true value. **E:** Shift ($\mathcal{S}$) calculated from the truncated distribution versus the approximated value of $\lambda$. **F:** $\mathcal{S}$ calculated from the truncated distribution versus the approximated value of $\gamma_1$. **G:** Approximated kurtosis ($\kappa$) vs. the true value.

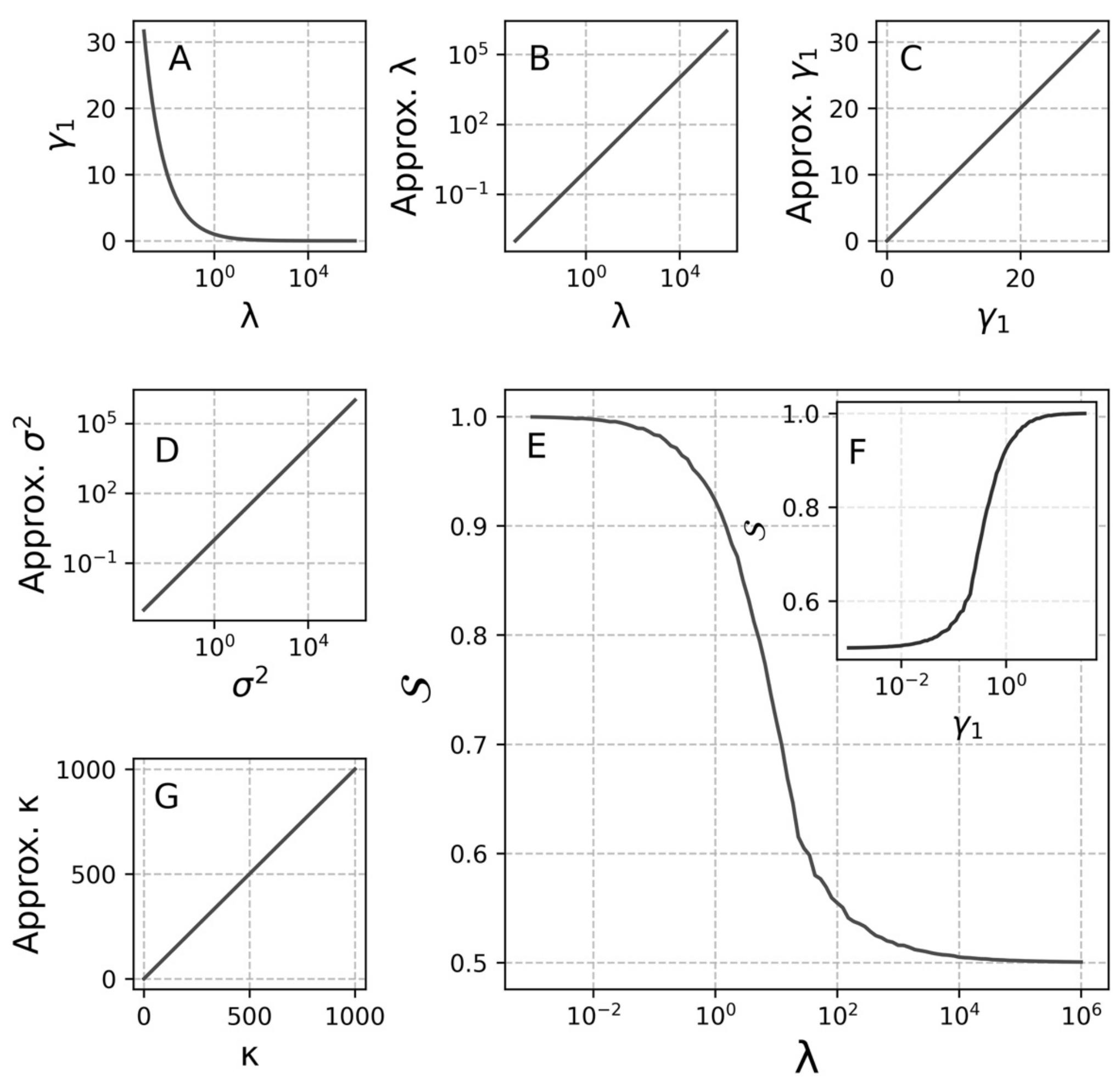

### 2.2 Relationship between shift and skewness

By definition, a relationship between $\mathcal{S}$ and skewness only exists within a bounded interval. When this condition is satisfied, shift and skewness may strongly covary, as demonstrated for a truncated, renormalized Poisson distribution (Fig 1). The relationship, $\mathcal{S} \rightarrow \frac{1}{2}$ as $\gamma_1 \rightarrow 0$, is also apparent for similarly truncated, renormalized versions of the Gamma, Negative Binomial, and Chi-squared distributions (Fig 2, A-C). Formal relationships between shift and skewness arise when $[L, U]$ coincides with the support of a naturally bounded distribution. Consider $\gamma_1$ and $\mathcal{S}$ for the Beta distribution (Eqn 2):

$$\mathcal{S} = \frac{\beta}{\alpha + \beta}, \qquad \gamma_1 = \frac{2(\beta - \alpha)\sqrt{\alpha + \beta + 1}}{(\alpha + \beta + 2)\sqrt{\alpha\beta}}.$$

Here, the asymptotic relationship between shift and skewness is evident:

$$\text{As } \mathcal{S} \rightarrow 1,\ \alpha \rightarrow 0 \text{ and } \sqrt{\alpha\beta} \rightarrow 0. \text{ Thus, } \gamma_1 \rightarrow +\infty.$$

$$\text{As } \mathcal{S} \rightarrow 0,\ \alpha \rightarrow +\infty \text{ and } 2(\beta - \alpha)\sqrt{\alpha + \beta + 1} \rightarrow -\infty. \text{ Thus, } \gamma_1 \rightarrow -\infty.$$

Due to $(\beta - \alpha)$ in the numerator of $\gamma_1$, if $\alpha = \beta$, then $\mathcal{S} = \frac{1}{2}$ and $\gamma = 0$. When plotted, the shape of the relationship is clear (Fig 2D). Similar relationships can be shown for the Binomial and Hypergeometric distributions (Fig 2E, F).

Outside of strict theoretical conditions, shift and skewness are not necessarily related. If the distribution $D$ is symmetric about the midrange of $[L, U]$, then $\gamma_1 = 0$ and $\mathcal{S} = \frac{1}{2}$, provided the central tendency $C$ equals $(L + U)/2$. However, $\mathcal{S} = \frac{1}{2}$ necessitates neither $\gamma = 0$ nor symmetry about the midrange of $[L, U]$. For example, let:

$$D(X) = [0.05, 0.2, 0.55, 0.1, 0.1], \quad X = [0, 1, 2, 3, 4],$$

Taking $L = 0$, $U = 4$, and $C = \bar{x} = (L + U)/2 = 2$ yields $\mathcal{S} = \frac{1}{2}$ and $\gamma_1 \approx 0.3514$.

**Figure 2.** Shift ($\mathcal{S}$) versus skewness ($\gamma_1$) for probability distributions across varying parameter values. **Top row:** Analyses based on truncated, renormalized distributions preserving the moments of the original unbounded forms. **Bottom row:** Analyses based on analytically derived measures of $\mathcal{S}$ for naturally bounded distributions.

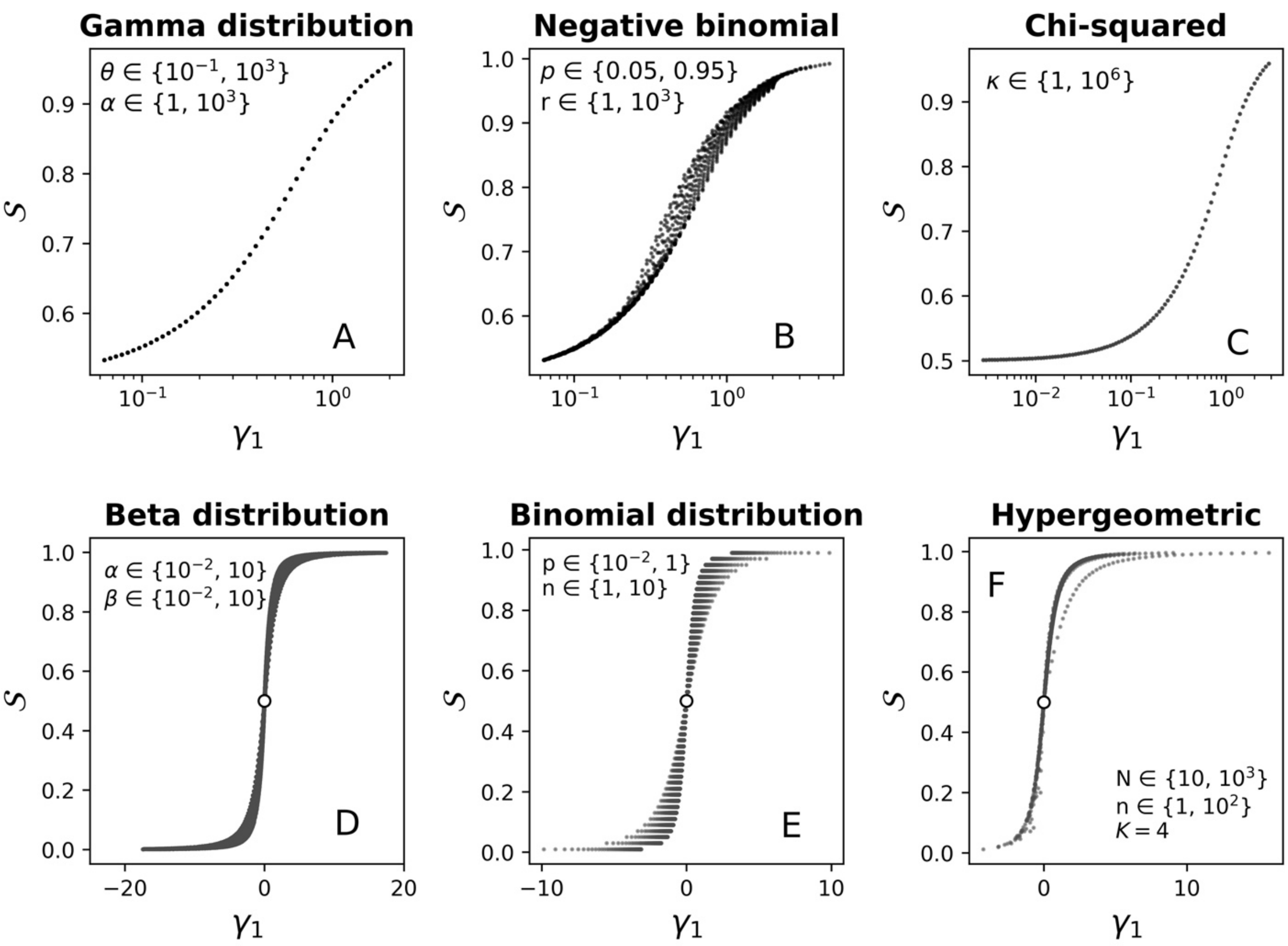

Shift may also vary independently of skewness. Recall that the interval [$L$, $U$] need not be the minimal interval containing $D$ but may contain regions of zero probability or zero frequency. Such may be the case when accounting for absences in sample distributions, when comparing a sample to a benchmark, or when comparing distributions on a common domain.

Consider the following cases where $D$ is a discrete frequency distribution defined on the domain $X = \{1, 2, 3, 4, 5\}$. Let $L = 1$, $U = 5$, and $C = \bar{x}$:

$D(X) = [0, 0, 0, 0, 4]$, $\bar{x} = 5$, $\gamma_1$ undefined, $\mathcal{S} = 0$,

$D(X) = [0, 0, 0, 2, 2]$, $\bar{x} = 4.5$, $\gamma_1 = 0$, $\mathcal{S} = 1/8$,

$D(X) = [0, 0, 1, 2, 1]$, $\bar{x} = 4$, $\gamma_1 = 0$, $\mathcal{S} = 1/4$,

$D(X) = [0, 1, 2, 1, 0]$, $\bar{x} = 3$, $\gamma_1 = 0$, $\mathcal{S} = 1/2$,

$D(X) = [1, 2, 1, 0, 0]$, $\bar{x} = 2$, $\gamma_1 = 0$, $\mathcal{S} = 3/4$,

$D(X) = [2, 2, 0, 0, 0]$, $\bar{x} = 1.5$, $\gamma_1 = 0$, $\mathcal{S} = 7/8$,

$D(X) = [4, 0, 0, 0, 0]$, $\bar{x} = 1$, $\gamma_1$ undefined, $\mathcal{S} = 1$.

While shift $\mathcal{S}$ reflects the location of the mean, $\gamma_1$ is either undefined or zero. These same results would also hold for other measures of skewness that do not account for zero-valued frequencies, such as Pearson's first and second skewness coefficients, Bowley's quartile-based measure of skewness, Kelly's decile-based measure of skewness, L-skewness, and the Groeneveld-Meeden coefficient of skewness.

Ultimately, the relationship between shift and skewness, whether direct, indirect, statistically related, or otherwise decoupled depends upon the existence of the finite interval $[L, U]$ and whether the interval contains regions of zero density or zero frequency, as well as the type of central tendency (e.g., mean, median, mode) represented by $C$.

## 3 Comparative measures of shift

The shift statistic $\mathcal{S}$ is positional, non-negative, and normalized. These properties allow direct comparisons of $\mathcal{S}$ between distributions that may or may not share a common domain, so long as $\mathcal{S}$ is based on the same form of central tendency, e.g., mean, median, mode. Below, we describe two comparative measures of shift.

### 3.1 $\Delta\mathcal{S}$ as a directional difference

Let $\mathcal{S}_1$ and $\mathcal{S}_2$ denote the shift values of two distributions $D_1$ and $D_2$, respectively, defined on finite intervals $[L_1, U_1]$ and $[L_2, U_2]$. Let $C_1$ and $C_2$ denote the corresponding central tendencies defined using the same measure, e.g., mean. Define the difference in shift as:

$$\Delta\mathcal{S} = \mathcal{S}_1 - \mathcal{S}_2, \text{ with } -1 \leq \Delta\mathcal{S} \leq 1. \tag{3}$$

The measure $\Delta\mathcal{S}$ expresses the direction and magnitude of difference between the positions of $C_1$ and $C_2$ relative to the upper bound $U$, as a proportion of the width of $[L, U]$:

If $[L_1, U_1] = [L_2, U_2] = [L, U]$:

- $\Delta\mathcal{S} = 0 \;\Rightarrow\;$ $C_1$ and $C_2$ lie in the same position within $[L, U]$.
- $\Delta\mathcal{S} < 0 \;\Rightarrow\;$ $C_2$ lies farther from $U$ than $C_1$, by $|\Delta\mathcal{S}| \times (U - L)$.
- $\Delta\mathcal{S} > \;\Rightarrow\;$ $C_1$ lies farther from $U$ than $C_2$, by $|\Delta\mathcal{S}| \times (U - L)$.

If $[L_1, U_1] \neq [L_2, U_2]$:

- $\Delta\mathcal{S} = 0 \;\Rightarrow\;$ $C_1$ and $C_2$ lie at equal relative positions.
- $\Delta\mathcal{S} < 0 \;\Rightarrow\;$ $C_2$ is relatively closer to $L_2$ than $C_1$ is to $L_1$ by a proportion $|\Delta\mathcal{S}|$.
- $\Delta\mathcal{S} > \;\Rightarrow\;$ $C_1$ is relatively closer to $L_1$ than $C_2$ is to $L_2$ by a proportion $|\Delta\mathcal{S}|$.

Alternatively, for a single distribution $D$ defined on $[L, U]$, let $C_1$ and $C_2$ represent two distinct central tendencies (e.g., mean and median). Then, $\Delta\mathcal{S}$ remains defined as above and interpreted within the common domain $[L, U]$.

### 3.2 $|\Delta\mathcal{S}|$ as a distance metric

Proper distance metrics must satisfy four requirements [15, 8]. Denote distance as $d$, these are:

1. Reflexivity: $d(D_1, D_1) = 0$
2. Non-negativity: $0 \leq d(D_1, D_2)$
3. Symmetry: $d(D_1, D_2) = d(D_2, D_1)$
4. Triangle inequality: $d(D_1, D_3) \leq d(D_1, D_2) + d(D_2, D_3)$

While $\Delta\mathcal{S}$ satisfies only reflexivity, its absolute value $|\Delta\mathcal{S}|$ satisfies all four properties of a proper distance metric, as a consequence of the properties of absolute differences:

Reflexivity: $|\mathcal{S}_1 - \mathcal{S}_1| \geq 0$ by definition
Non-negativity: $|\mathcal{S}_1 - \mathcal{S}_1| \geq 0$ by definition
Symmetry: $|\Delta\mathcal{S}| = |\mathcal{S}_1 - \mathcal{S}_2| = |\mathcal{S}_2 - \mathcal{S}_1|$
Triangle inequality: $|\mathcal{S}_1 - \mathcal{S}_3| \leq |\mathcal{S}_1 - \mathcal{S}_2| + |\mathcal{S}_2 - \mathcal{S}_3|$, by the inequality for absolute values

Thus, while $\Delta\mathcal{S}$ is a signed (directional) measure of proportional difference, its absolute value $|\Delta\mathcal{S}|$ satisfies the properties of a proper distance metric.

### 3.3 Relationships of $|\Delta\mathcal{S}|$ to established distance metrics

We examined how $|\Delta\mathcal{S}|$ relates to distance metrics that conceptually relate to shift: Wasserstein Distance, Energy Distance, Cramér Von Mises Distance, Hellinger Distance, Total Variation Distance, and Kolmogorov-Smirnov distance (Table 2) [15, 31, 27, 3, 20, 4]. We computed Wasserstein Distance and Energy Distance using the Statistical Functions module of the python-based scientific computing library, SciPy [54]. Calculations of other distance metrics were based on Deza and Deza (2014) [15].

We computed $|\Delta\mathcal{S}|$ and other distance metrics on $10^3$ pairs of randomly parameterized Beta-distributed probability density functions. Within each pair, $\alpha$ and $\beta$ were selected at random on a logarithmic scale from $10^{-1}$ to $10^1$. Additionally, $\alpha$ and $\beta$ were selected independently for each distribution within each pair. This approach allowed a wide exploration of parameter space and high potential for comparisons between distributions that ranged from highly similar in shape to highly disparate. We then characterized relationships between $|\Delta\mathcal{S}|$ and each other distant metric using ordinary least-squares regression.

Values of $|\Delta \mathcal{S}|$ were most strongly related to those of Wasserstein Distance (slope = 1.03, $r^2$ = 0.99) (Fig 3). The strength of this nearly 1-to-1 relationship was surprising given the simplicity of $|\Delta \mathcal{S}|$ relative to the optimal transport theory that underpins Wasserstein Distance (Table 2). Values of $|\Delta \mathcal{S}|$ were also strongly related to those of Energy Distance (slope = 0.73, $r^2$ = 0.97) and Cramér Von Mises Distance (slope = 0.97, $r^2$ = 0.98) (Fig 3). Values of $|\Delta \mathcal{S}|$ were less strongly related to values of Hellinger Distance, Total Variation Distance, and Kolmogorov-Smirnov distance (Fig 3).

**Figure 3.** Ordinary least squares regression of $|\Delta \mathcal{S}|$ on other measures of distance or divergence; *m* is the slope of the relationship. Here, $|\Delta \mathcal{S}|$ is calculated based on the mean of arithmetic values

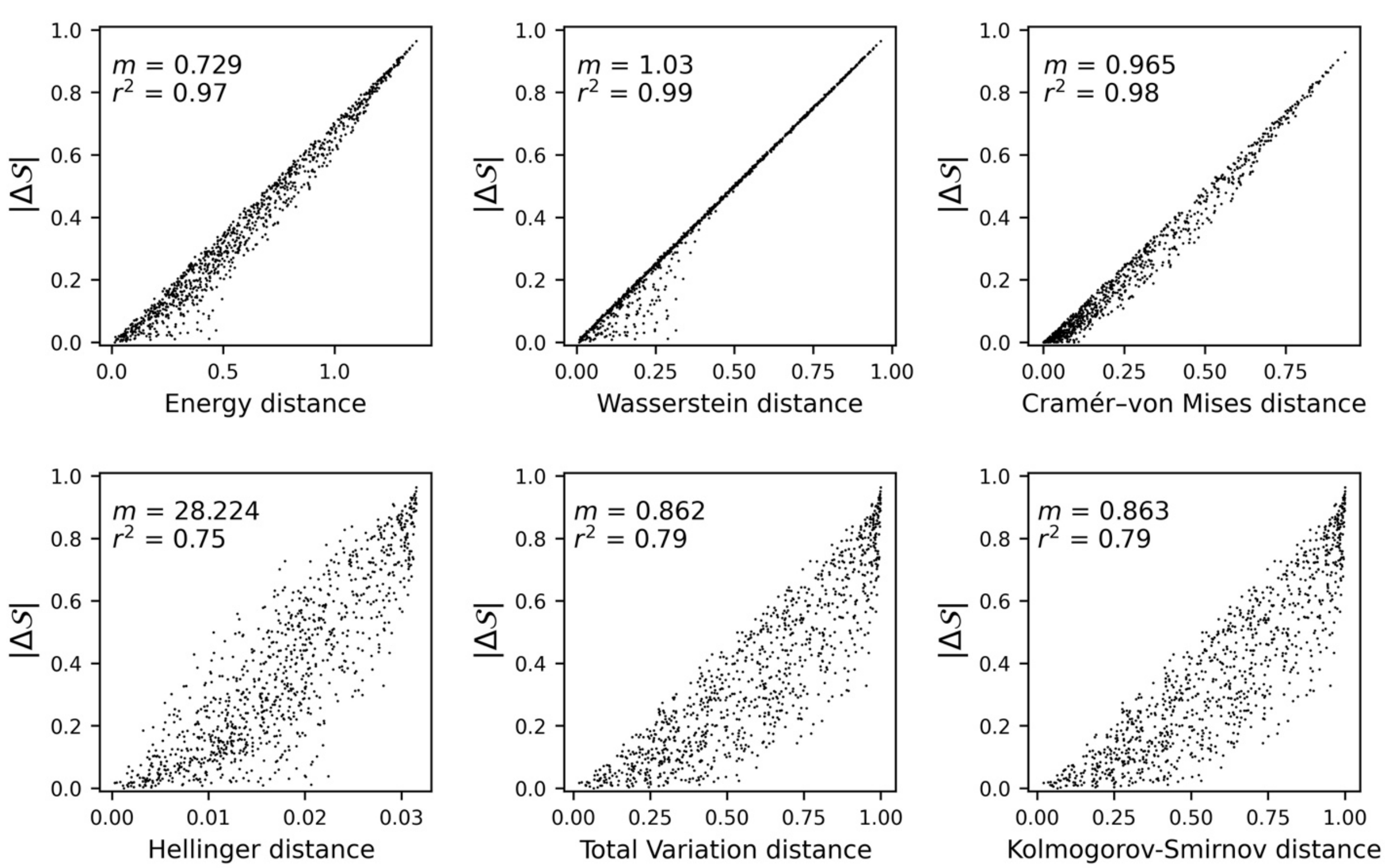

The relationship between $|\Delta S|$ and Wasserstein distance was notable given that $|\Delta S|$ relies on summary statistics while Wasserstein distance is based on optimal transport and is often used in generative artificial intelligence [10, 22, 31]. As implemented in SciPy, Wasserstein distance for one-dimensional distributions is computed from the area between two marginal cumulative distributions, avoiding use of optimal transport but requiring pairwise comparisons on sorted samples; computational bottlenecks that $|\Delta S|$ avoids. Comparing the algorithm of SciPy to our implementation of $|\Delta S|$, we calculated runtimes across uniform random samples of size $10^2$ to $10^6$, with $10^4$ replicates each, using a 2019 MacBook Pro (2.4 GHz Quad-Core Intel Core i5, 8 GB 2133 MHz LPDDR3 RAM).

At samples sizes of $10^2$, the mean compute time in milliseconds for $|\Delta S|$ was nearly 1.5 times faster than the SciPy implementation of Wasserstein Distance (Table 3). The difference between the two increased over orders of magnitude where, at sizes of $10^6$, the mean compute time for $|\Delta S|$ was nearly 270 times faster than the SciPy implementation of Wasserstein Distance. If the relationship between $|\Delta S|$ and Wasserstein Distance generally holds, then $|\Delta S|$ may provide a more efficient, scalable, and intuitive measure.

**Table 3.** Average computing times in milliseconds (ms) for $|\Delta S|$ and Wasserstein Distance across $10^4$ replicates of each sample size ($n$). SD = standard deviation

| ***n*** | **$|\Delta S|$ (ms) ± SD** | **Wasserstein (ms) ± SD** | **Wasserstein / $|\Delta S|$ ± SD** |
|---|---|---|---|
| $10^2$ | 0.040 ± 0.015 | 0.059 ± 0.021 | 1.496 ± 0.291 |
| $10^3$ | 0.047 ± 0.016 | 0.350 ± 0.051 | 7.635 ± 0.866 |
| $10^4$ | 0.078 ± 0.035 | 4.214 ± 0.537 | 57.032 ± 9.900 |
| $10^5$ | 0.298 ± 0.070 | 53.615 ± 3.655 | 184.255 ± 22.003 |
| $10^6$ | 2.704 ± 0.455 | 738.961 ± 64.907 | 276.610 ± 27.097 |

## 4 Applications

### 4.1 Shift as a measure of species rarity

Distributions of abundance among species are intensively studied in the field of ecology and are often interpreted in terms of ecological processes [35, 37-39, 41, 51]. However, the number of sampled individuals ($N$) and the number of species ($S$) among them impose numerical constraints on the shapes of abundance distributions, obscuring ecological interpretations with numerical artifacts [35]. For example, studies have reported that species rarity scales with $N$ as:

$$R_{\gamma} \propto z \cdot \log_{10}(N), \quad 0.1 \leq z \leq 0.2,$$

where, $R_{\gamma}$ is a log-modulo transformation of skewness:

$$R_{\gamma} = \log(|\ \gamma_1\ | + 1), \text{ if } 0 \leq \gamma_1,$$
$$R_{\gamma} = -\log(|\ \gamma_1\ | + 1), \text{ if } \gamma_1 < 0.$$

However, a positive relationship between $\gamma_1$ and $N$ is expected based on how $N$ and $S$ constrain the shape of the abundance distribution [35]. Considering shift as a measure of rarity begs the question of whether $\mathcal{S}$ is any less affected.

For this analysis, we used the same methodology that was used to show how $N$ and $S$ constrain the shape of species abundance distributions [35]. This approach uses combinatorial algorithms to draw uniform random samples of integer partitions satisfying $N$ and $S$, where each partition is an unordered way of summing $S$ positive integers to obtain $N$ [35]. Using algorithms of [34], we generated $10^3$ partitions for values of $N \in [10, 10^4]$ and values of $S$ drawn at random using an expected scaling relationship: $S = 1.75 \cdot N^b$, where $0.25 \leq b \leq 0.5$ and $b$ was a uniform random variate [33]. For each partition, we calculated shift $\mathcal{S}$ on $\log_{10}$-transformed abundances, taking $L$ as the minimum, $U$ as the maximum, and $C$ as the mean. We also computed $R_{\gamma}$ and the percentage of species represented by a single individual, $R_1$.

We found that the rate at which $R_\gamma$ has been reported to scale with $N$ (i.e., 0.1 to 0.2) can largely be explained by how total abundance $N$ and species richness $S$ numerically constrain variation in $R_\gamma$, where $R_\gamma$ scaled with $N$ at a rate of 0.1 (Fig 4). Hence, there may be little information in the previously reported scaling law regarding the role of ecological processes and mechanisms in driving the relationship between rarity and total abundance $N$. Surprisingly, $R_\gamma$ scaled even more strongly with $S$ than it did with $N$. This too was merely a consequence of numerical constraints.

**Figure 4. Top row:** Species rarity versus $\log_{10}$-transformed sample abundance ($N$). **Bottom row**: Species rarity versus number of species ($S$). Each data point is one of $10^3$ random samples from integer partition based feasible sets. **Left:** Species rarity as log-modulo skewness, with $z$ representing the scaling exponent. **Center:** Percentages of species represented by 1 individual. **Right**: Shift, $\mathcal{S}$, calculated from log-transformed abundances and their mean, with $m$ representing the slope of the relationship.

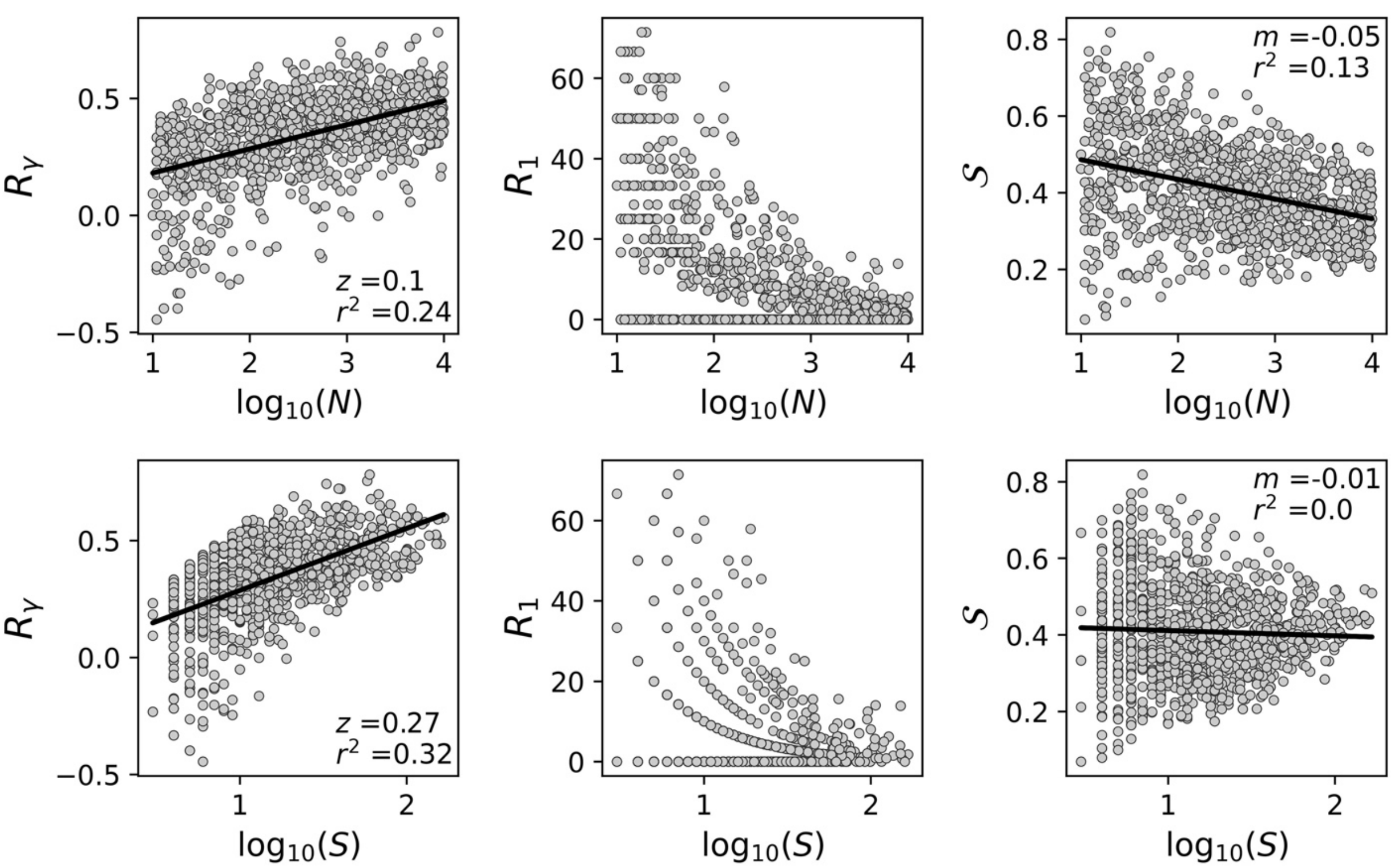

While $R_\gamma$ tended to increase with $N$ and $S$, the percentage of species represented by a single individual, $R_1$, strongly decreased (Fig 4). This result is reasonable, considering that greater average species abundance ($N/S$) reduces the likelihood of singletons. Consequently, $R_1$ is a poor characterization of rarity. As suggested earlier, we expect similar artifacts to emerge among other threshold based measure of rarity. By comparison to $R_\gamma$ and $R1$, rarity based on shift trended weakly with $N$ and did not trend at all with $S$ (Fig 4). In this way, rarity based on shift appears to be a comparatively independent measure of the concentration of species toward low abundances, particularly in its relation to the number of species.

**4.2 Shift as a measure of poverty**

As previously mentioned, measures of poverty are invariably based on arbitrary and often controversial poverty lines (Table 1) [2, 24, 55]. Though poverty lines may serve a practical purpose, the question begs to be answered whether poverty can be measured without relying on poverty lines but in a way that does not produce disparate with established poverty measures. Here, we demonstrate shift, $\mathcal{S}$, as such a measure.

We used $10^3$ simulated income distributions among families of varying size. In each simulation, we generated a distribution of household sizes for $10^3$ households by sampling from a Poisson distribution with a mean ($\lambda$) of 2.5 corresponding to an expected household size of 2.5 [26]. We then simulated a distribution of household incomes using a lognormal distribution, with the location parameter ($\mu$) of 11 and a scale parameter ($\sigma$) treated as a random variate, $1 \leq \sigma \leq 3$. This approach produced a median income (80,907 ± 6,271 SD) similar to that of the 2023 U.S. median of $80,610 [23]. We then used the 2025 U.S. Federal Poverty Guidelines for households of different sizes to determine the poverty status of each household (see: https://aspe.hhs.gov/). Using the resulting data, we calculated the Gini coefficient to assess overall income inequality

[24], and calculated standard measures of poverty: Headcount Ratio, Poverty Gap Index, Sen Index (Table 1). We then performed ordinary least-squares regressions between each of the above measures and $\mathcal{S}$, where $\mathcal{S}$ was based on $\log_{10}$-transformed values.

$\mathcal{S}$ was strongly related to each measure of poverty and the Gini index of inequality (Fig 5). With respect to our particular model, $\mathcal{S}$ explained 86% to 88% of variation among poverty measures (Fig 5A-C). Though not linearly related to the Gini index, $\mathcal{S}$ explained 95% of variation in the Gini index when using a logistic relationship and calculating the coefficient of determination using observed vs. predicted values, i.e., slope of 1 and *y*-intercept of 0 (Fig 5D). While established measures of poverty reflect the concentration of a population towards its lowest incomes, the ability to measure this concentration without poverty lines is encouraging.

**Figure 5.** Relationships of measures of poverty and inequality to shift, $\mathcal{S}$, where $\mathcal{S}$ was calculated from log-transformed household incomes and their mean. Within each plot, each point is an outcome of an independent simulation

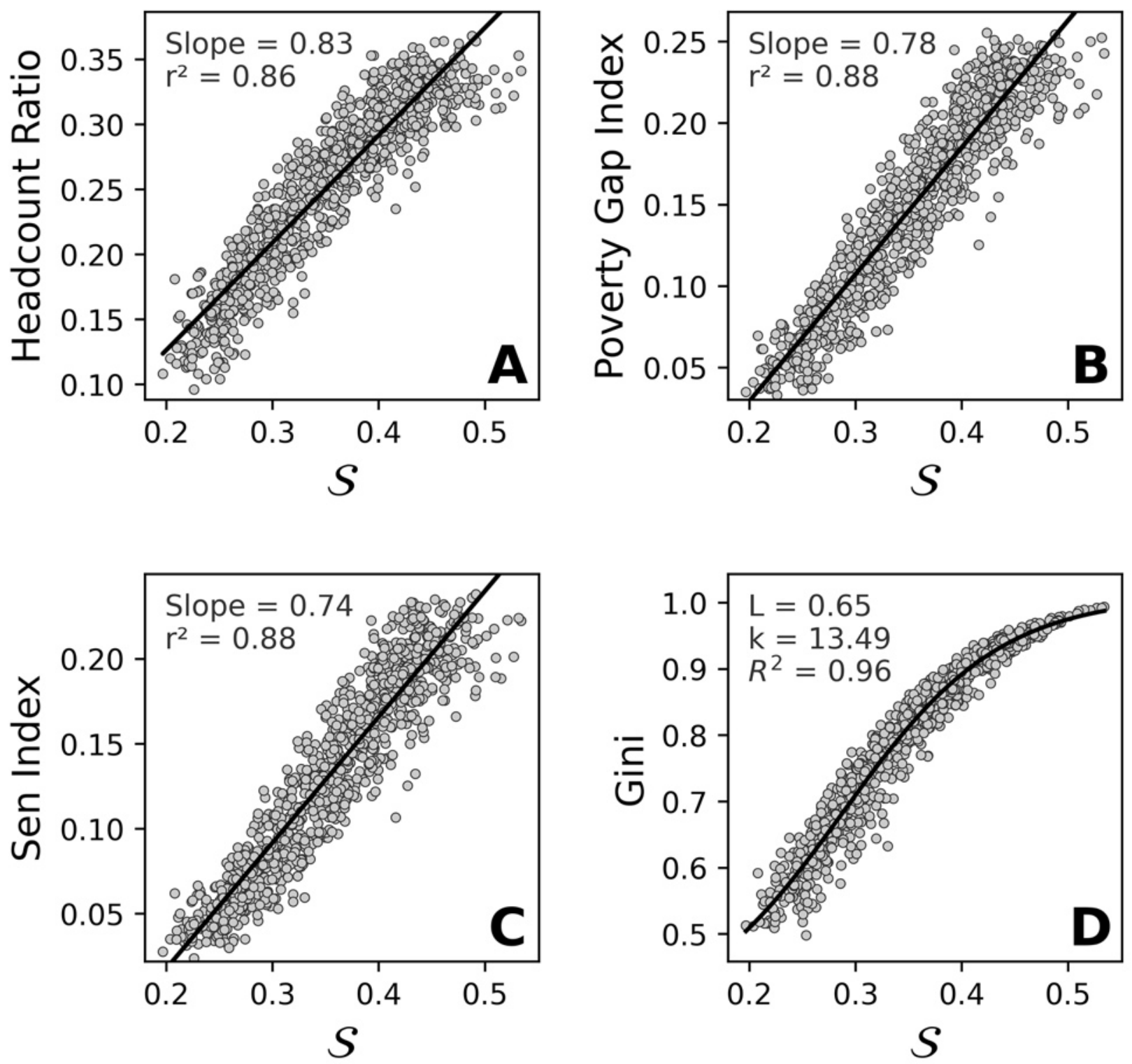

### 4.3 Application of shift to image analysis

Distance metrics are often used to compare histograms of image data [7, 11, 14, 18, 27]. As already demonstrated, values of $|\Delta\mathcal{S}|$ strongly relate to those of Wasserstein distance, suggesting the two may perform similarly in image analytical tasks. However together, $\mathcal{S}$, $\Delta\mathcal{S}$, and $|\Delta\mathcal{S}|$ contain information that Wasserstein distance does not.

We used the OpenCV (v4.9.0) computer vision library to construct a MP4-formatted video of 100 frames (resolution = 300 × 300 pixels) [6]. Hues (i.e., color types) were represented in HSV (Hue, Saturation, Value) color space [49]. Within each frame, a stationary disk with a radius of 100 pixels was centered against black background. As the video played, the disk transitioned from hues of red to orange, yellow, green, blue, magenta, and back to red for a total of two HSV hue cycles. To simulate discrete events, the disk was omitted at frames 9, 19, 29, and 39. For analysis, we used standard conversion techniques of the OpenCV library to convert each video frame to a one-dimensional histogram of 256 bins representing different light intensity values [6, 11]. Using these histograms, we computed $\mathcal{S}$ for each frame and computed $\Delta\mathcal{S}$, $|\Delta\mathcal{S}|$, and Wasserstein distance for each frame relative to the first, producing a signal of change from a point of origin.

Across frames, signals produced from $|\Delta\mathcal{S}|$ and Wasserstein distance were indiscernible (Fig 6). These metrics equally characterized how light intensity within video frames changed across two cycles of hue and both equally captured the disappearance of the disk. However, the values produced by the two metrics differed (Fig 6). While $|\Delta\mathcal{S}|$ ranges from 0 to 1, Wasserstein distance is determined by the scale and range of values in the distributions. The signal produced by $\mathcal{S}$ was indiscernible from that produced by $\Delta\mathcal{S}$ but this was a consequence of having computed $\Delta\mathcal{S}$ relative to the first frame.

**Figure 6. Top:** Select video frames with frame numbers indicated above. **Center:** The graph of Wasserstein Distance maps directly to the graph of $|\Delta\mathcal{S}|$; hence, there appears to be only one line. **Bottom:** The graph of $\Delta\mathcal{S}$ maps directly to the graph of $\mathcal{S}$ because the first frame is the reference point for each measure of $\Delta\mathcal{S}$. Measures of $\Delta\mathcal{S}$ were based on the mean of arithmetic values

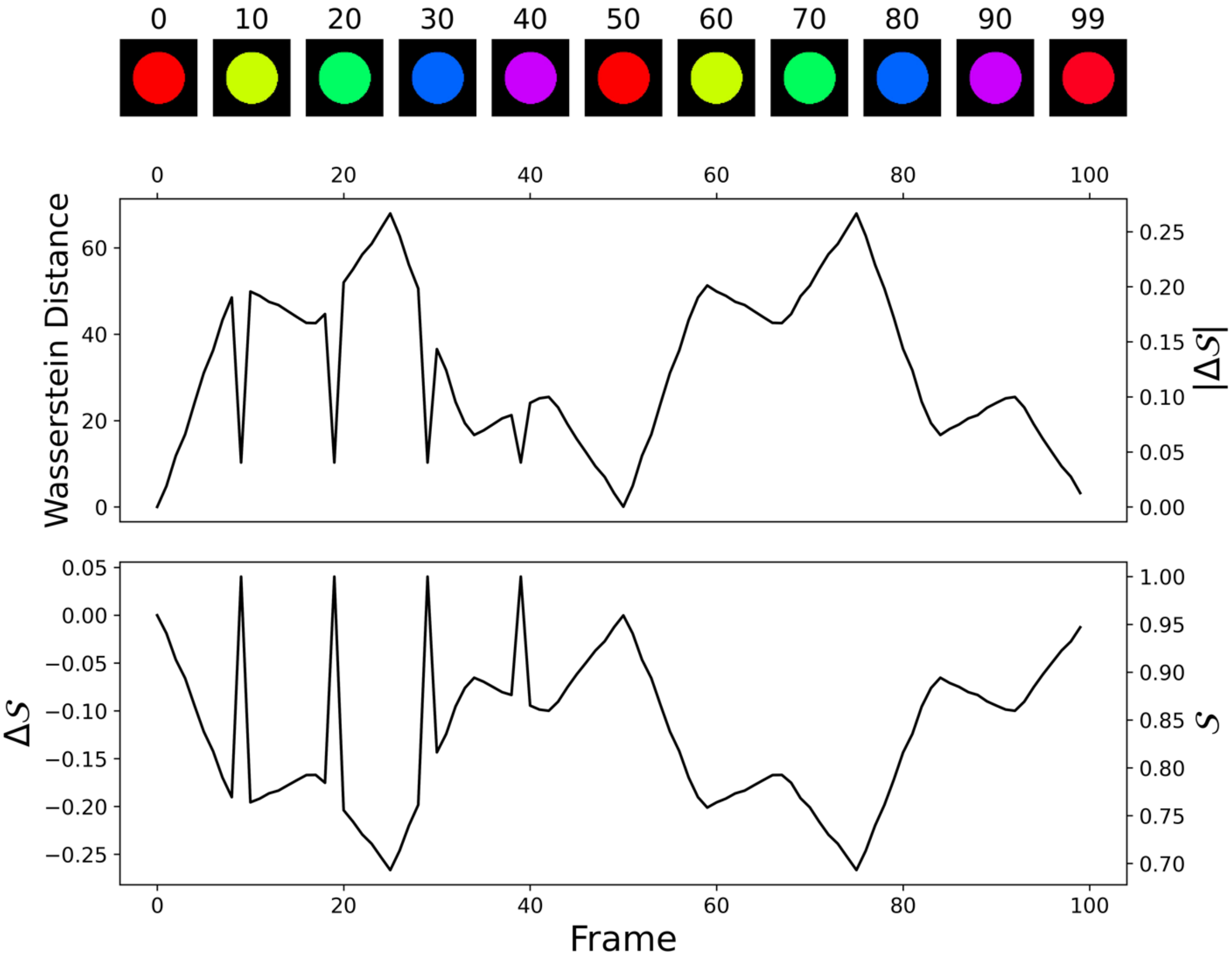

This exercise reinforced the relationship between $|\Delta\mathcal{S}|$ and Wasserstein distance and revealed that meaningful information can be gained from the combined used of $\mathcal{S}$, $\Delta\mathcal{S}$, and $|\Delta\mathcal{S}|$. For example, spikes indicating the disappearance of the disk were largest for $\Delta\mathcal{S}$, due to its values being distributed across zero; in contrast to the absolute values of distance measures (Fig 6). Put simply, losing the sign of a difference obscures the size of a difference. Beyond its associated measures of $\Delta\mathcal{S}$ and $|\Delta\mathcal{S}|$, $\mathcal{S}$ contains contextual information. For example, the initial value of $\mathcal{S} \approx 0.95$ indicated initially low overall light intensity (i.e., a red disk against a black background), while spikes in $\mathcal{S}$ to the lower bound of intensity revealed that the image went black when the disk disappeared.

Taken together, $\mathcal{S}$ indicates the location of a central tendency relative to a bound, $\Delta\mathcal{S}$ indicates the difference in $\mathcal{S}$ with respect to direction and a point of reference, and $|\Delta\mathcal{S}|$ provides an unambiguous absolute distance that may be related to the concepts of optimal transport that underpin Wasserstein Distance.

## 5 Concluding Remarks

The concentration of a distribution toward a lower (or upper) bound is a conceptually simple property that has previously lacked a general, global descriptive statistic. We referred to this property as shift and based its measurement on the distance from a central tendency to an upper bound, expressed as a proportion of a finite interval. Though not defined for distributions lacking natural or imposed bounds, $\mathcal{S}$ an otherwise flexible statistic. Measurements of $\mathcal{S}$ can be based on any well-defined central tendency and can be analytically derived or computed for discrete and continuous distributions of probabilities or frequencies, and implicitly account for regions of zero probability or frequency. As demonstrated, $\mathcal{S}$ naturally extends to a directional measure of difference $\Delta\mathcal{S}$ and to a proper distance metric $|\Delta\mathcal{S}|$.

Despite its simplicity, $|\Delta\mathcal{S}|$ strongly relates to measures of greater complexity such as Energy distance, Cramér Von Mises distance, and Wasserstein distance. The latter of these is central to Wasserstein Generative Adversarial Networks (WGANs), which underpin models of generative artificial intelligence [1, 22, 31]. Our preliminary comparison of the time taken to compute $|\Delta\mathcal{S}|$ versus that of an efficient algorithm for Wasserstein distance revealed that compute times for $|\Delta\mathcal{S}|$ at sample sizes of $10^6$ were lower than compute times for Wasserstein distance at sample sizes of $10^3$. Assuming this improved speed coincides with similar performance, the implications may be profound for computation intensive applications that rely on Wasserstein distance. In particular, Wasserstein distance is often computed over *n*-dimensional distributions, which excludes the algorithm used in this study and instead requires solving optimal transport via linear programmatic computation of cost matrices.

Beyond its use as a comparative statistic, the concept of shift closely relates to a common pattern in nature, i.e., uneven distributions of wealth and abundance. These patterns are almost universally heavy-tailed, right-skewed, and highly shifted toward a lower bound [41]. Relatedly, the concept of shift closely relates to concepts of scarcity and diminutiveness, a property of size. In particular, the positive skew of body size distributions is often used as evidence of ecological constraints and evolutionary trends towards smaller body sizes [43]. Moreover, the shift statistics presented here are no less relevant to the analysis of left-skewed distributions. In particular, distributions of log-transformed species abundances are often left-skewed, provoking greater explanations for the excess of rare species [38, 39, 51]. Distributions of human lifespan and cognitive scores are also highly left-skewed, revealing shifts towards upper bounds of longevity and cognition [56, 57].

The concentration of a distribution toward a lower bound is often characterized with measures that may suffer from critical statistical artifacts and arbitrary thresholds, or measures of asymmetry that coincidentally or conditionally relate to the location of concentration. Though measures based on poverty lines and rarity thresholds will undoubtedly remain in popular use and have proven valuable, shift offers a flexible global measure that is similar in generality to the Gini index. The shift statistic $\mathcal{S}$ is simpler in calculation than several established poverty indices (Table 1), more intuitive than unbounded transformed values of skewness, and extends to a directional measure of difference $\Delta\mathcal{S}$ and its associated distance metric $|\Delta\mathcal{S}|$.

Importantly, our work has limitations. First, the shift statistic $\mathcal{S}$ is incalculable for unbounded distributions. While some theoretical distributions can be truncated and renormalized to preserve the moments of the unbounded form, doing so produces a fundamentally object. Second, despite their potential breadth of use and advantages, our shift statistics are based on central tendencies that capture relatively little information about a given distribution. Third, we did not explore the calculation of $\mathcal{S}$ for distributions of ordinal categories or for varying regimes of histogram binning. Fourth, our simulations and computational time comparisons were largely exploratory and were not intended as focused studies. Finally, there may be consequential properties of $\mathcal{S}$, $\Delta\mathcal{S}$, and $|\Delta\mathcal{S}|$ that we did not explore or acknowledge.

In summary, the shift statistics presented here quantify a simple but overlooked statistical property that may be of value across fields of study and diverse applications whether as complements to established statistics, alternatives to less intuitive indices or computationally demanding metrics, or as a standalone suite of targeted measures. Given their simplicity, we suspect that improvements to these statistics, if not, entirely new, sophisticated, and robust alternatives will follow.

**Acknowledgments**

The authors thank Dr.'s Ethan P. White and Xiao Xiao for fruitful discussions.

**Data availability statement**

Data and source code needed to reproduce the findings and figures herein are available in a public GitHub repository: https://github.com/Rush-Quality-Analytics/distributional_shift.

**Disclosures and declarations**

This study was not supported by external funding and the authors declare no conflict of interests.